\documentclass[prl,twocolumn,preprintnumbers,amsmath,amssymb]{revtex4}
\usepackage{dcolumn}
\usepackage{bm}
\usepackage{graphicx}
\begin{document}

\title{Universality in Oscillating Flows} \date{\today}

\author{K. L. Ekinci\footnote{Corresponding author: ekinci@bu.edu}}
\affiliation{Department of Aerospace and Mechanical Engineering, Boston University, Boston, Massachusetts, 02215}
\author{D. M. Karabacak\footnote{Current address: Holst Centre/IMEC-NL, Eindhoven, The Netherlands.}}
\affiliation{Department of Aerospace and Mechanical Engineering, Boston University, Boston, Massachusetts, 02215}
\author{V. Yakhot}
\affiliation{Department of Aerospace and Mechanical Engineering, Boston University, Boston, Massachusetts, 02215}

\begin{abstract}
We show that oscillating flow of a simple fluid in both the Newtonian and the non-Newtonian regime can be described by
a universal function of a single dimensionless scaling parameter $\omega\tau$, where $\omega$ is the oscillation
(angular) frequency and $\tau$ is the fluid relaxation-time; geometry and linear dimension bear no effect on the flow.
Experimental energy dissipation data of mechanical resonators in a rarefied gas follow this universality closely in a
broad linear dimension ($10^{-6}$ m$< L < 10^{-2}$ m) and frequency ($10^5$ Hz $< \omega/2\pi < 10^8$ Hz) range. Our
results suggest a deep connection between flows of simple and complex fluids.

\end{abstract}
\maketitle

The law of similarity \cite{reynolds1883,landau1959} is the most basic consequence of the Navier-Stokes equations of
fluid dynamics: two geometrically similar flows with different velocities and viscosities but equal Reynolds numbers
can be made identical by rescaling the measurement units. Similarity turns into universality \cite{tanaka1996} if
additional characteristics of the flow -- such as the constitutive stress-strain relations, the flow geometry, and
other dynamical aspects -- can all be incorporated into the scaling. For  steady flows past solid bodies, for example,
the law of similarity can be formulated \cite{landau1959} in terms of the velocity field $\textit{\textbf{u}}$:
\begin{equation}\label{u}
\textit{\textbf{u}} = \textit{\textbf{U}}f(\frac{\textit{\textbf{r}}}{L},{\mathop{\rm Re}} ).
\end{equation}
Here, $\textit{\textbf{U}}$ is a characteristic velocity of the flow, $\textit{\textbf{r}}$ represents the position vector, $L$ is a dynamically relevant linear dimension, and ${\mathop{\rm Re}\nolimits}  = UL/\nu$ is the Reynolds number with kinematic viscosity $\nu$. The
dimensionless scaling function $f$, which reflects subtle features of the flow, is found from the Navier-Stokes
equations. Defining the mean free-path and relaxation time as $\lambda$ and $\tau$, respectively, one can express $\nu$
in terms of these microscopic parameters as $\nu \approx \lambda ^2 /\tau  \approx \lambda c_s $, where $c_s$ is speed
of sound.

The equations of fluid dynamics can be formulated in the most general form as
\begin{equation}\label{fluidic}
\frac{\partial u_i }{\partial t} + (\textit{\textbf{u}} \cdot \nabla ){\rm{ }}u_i  = \nabla_j \sigma _{ij},
\end{equation}
with $i,j = x,y,z$. The stress tensor $\sigma _{ij}$ can be expanded in powers of the Knudsen number $({\rm{Kn}} =
\lambda /L)$ and the Weissenberg number $({\rm{Wi}} = \tau /T)$ \cite{landau1981,chen2004,cercignani1975}: $\sigma
_{ij}  = \sigma _{ij}^{(1)}  + \sigma _{ij}^{(2)}  + ...$. Newtonian fluid dynamics, where $\sigma _{ij} = \frac{\mu
}{2}(\frac{{\partial u_i }}{{\partial x_j }} + \frac{{\partial u_j }}{{\partial x_i }})$ with $\mu  = \rho \nu$,
corresponds to the first term of the expansion.  To see this, simply consider a steady flow past a bluff body of linear
dimension $L$:  $\frac{\sigma }{{\rho U^2 }} \approx \frac{\lambda }{L}\frac{{c_s }}{U} \approx
\frac{\rm{Kn}}{\rm{Ma}}$ given that the velocity gradient $\sim \frac{U}{L}$; here, ${\rm{Ma}} = U/c_s$ is the Mach
number. Thus, the Newtonian approximation is only valid when $\rm{Kn} \ll 1$. With $\lambda  \approx c_s \tau$, it is
easy to see that $\rm{Kn} = \frac{\lambda }{\textit{L}} \approx \frac{{\textit{c}_\textit{s}}}{\textit{U}}\frac{\tau
}{\textit{T}} \approx \frac{\rm{{Wi}}}{\rm{Ma}}$ and $\frac{\sigma }{{\rho U^2 }} \approx
\frac{{\rm{Wi}}}{{\rm{Ma^2}}}$. Consequently, $\rm{Kn} \ll 1$ and  $\rm{Wi} \ll 1$ correspond to the same physical
limit for fixed $\rm{Ma}$. This statement can also be extended to an unsteady flow past a bluff body, where the shed
vortices may be thought to introduce independent time ($\rm{Wi}$) and length scales ($\rm{Kn}$). Yet, the
characteristic oscillation frequency for vortex shedding is given by $\Omega  = \frac{{2\pi U \rm{St}}}{L}$, where
${\rm{St}} = \frac{L}{{UT}}$ is the Strouhal number. It is well-known that ${\rm{0}}{\rm{.1}} \le {\rm{St}} \le
{\rm{1}}$ over a wide range of $\rm{Re}$ depending on geometry. Thus, $\Omega \tau = \frac{{2\pi \tau
}}{T}\sim\frac{\lambda }{L}\frac{U}{{c_s }}$, and $\rm{Wi}$ and $\rm{Kn}$ are linked. It is then straightforward to
conclude that Newtonian fluid dynamics and emerging similarity relations, e.g. Eq. (\ref{u}), should be valid for {\it
slowly-varying large-scale} flows: $\rm{Kn} \ll 1$ and $\rm{Wi} \ll 1$.

Recent advances in micro- \cite{squires2005} and nanotechnology \cite{karabacak2007,verbridge2008}, biofluid mechanics
\cite{Stroock2002}, rheology \cite{Bird1987}, and so on have resulted in flows at unusual time ($\rm{Wi} \geq 1$) and
length  ($\rm{Kn} \geq 1$) scales, where Newtonian approximation breaks down. Many interesting phenomena, such as
elastic turbulence \cite{groisman2000}, structural relaxation of soft matter \cite{wyss2007} and enhanced heat transfer
in nanoparticle-seeded fluids \cite{wang1999}, have been observed in this range of parameters. Thus, a universal
description of Newtonian and non-Newtonian flow regimes is of great importance for both fundamental physics and
engineering.

\begin{figure*}
\includegraphics[trim=0mm 0mm 0mm 60mm,clip=true,scale=.37]{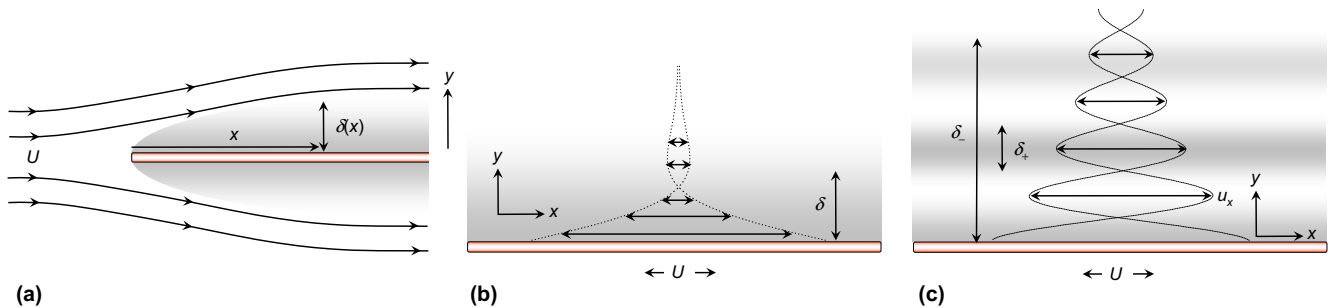}
\caption{(color online). Flow geometries and dynamical parameters. (a) Laminar steady flow over a semi-infinite flat
plate with a viscous boundary layer of thickness $\delta (x)$ and a free stream velocity $U$. (b) Unsteady flow
generated by an infinite plate oscillating at (angular) frequency $\omega$ with peak velocity $U$ in the Newtonian
limit ($\omega \tau \ll 1$). (c) The same problem in the non-Newtonian limit ($\omega \tau  \ge 1$). A wavelength
$\delta _ +  $ and a penetration depth $\delta _ -$ emerge instead of the viscous boundary layer thickness $\delta$.}
\label{schematic}
\end{figure*}

The similarity law given in Eq. (\ref{u}) is valid in the Newtonian regime only. For this regime, since both $\rm{Wi}$
and $\rm{Kn}$ are small,  $\rm{Re}$ is the only relevant dimensionless parameter. In the non-Newtonian regime, the
situation is different. For a similarity law valid in both Newtonian and non-Newtonian regimes, the scaling relation in
Eq. (\ref{u}) could be modified as
\begin{equation}\label{u_mod}
\textit{\textbf{u}} = \textit{\textbf{U}}f(\frac{\textit{\textbf{r}}}{L},\frac{\textit{\textbf{r}}}{\delta
},\frac{\lambda }{L},\frac{\lambda }{\delta },\frac{\tau }{T},{\mathop{\rm Re}\nolimits} ,{\rm{St}}).
\end{equation}
This is a {\it prescribed} relation including all relevant dimensionless parameters. In addition to the above-mentioned
parameters, a dynamic length scale $\delta$ (to be clarified below) may enter the scaling function $f$. The final form
of the scaling relation is determined by the physical nature of the flow. In order to assess the relative importance of
various dimensionless parameters and reduce the scaling relation in Eq. (\ref{u_mod}), we investigate the second order
correction to the stress tensor derived from the Boltzmann equation \cite{landau1981,chen2004,cercignani1975}:
\begin{equation}\label{stresstensor}
\sigma _{ij}^{(2)}  \approx \rho \lambda ^2 \frac{{\partial u_i }}{{\partial x_\alpha  }}\frac{{\partial u_j
}}{{\partial x_\alpha  }}.
\end{equation}

Considering first the shed vortices behind a body, we readily estimate that $\sigma ^{(2)} \approx \rho \lambda ^2 U^2
/L^2$. Thus, $\sigma ^{(1)}  + \sigma ^{(2)}  \approx \rho \nu\frac{U}{L} + \rho \lambda ^2 \frac{{U^2 }}{{L^2 }}
\approx \rho \nu\frac{U}{L}(1 + \frac{U}{{c_s }}\frac{\lambda }{L}) \approx \rho \nu\frac{U}{L}(1 + \frac{\tau }{T})$.
In this simple case, $\sigma$ appears as an expansion in powers of $\rm{Kn}$ or, equivalently, $\rm{Wi}$.

In another classic problem -- laminar steady flow over a semi-infinite flat plate illustrated in Fig.
\ref{schematic}(a) --  the relationship between $\rm{Kn}$ and $\rm{Wi}$ can still be stated, albeit with important
differences from above. The key dynamic feature of this flow is the viscosity-dominated boundary layer of thickness
$\delta  = \sqrt {\frac{{x \nu}}{U}} $. Here, $U$ is the free-stream velocity outside the boundary layer.  The
Newtonian approximation results in $\sigma _{xy}^{(1)}  = \rho \nu\frac{{\partial u_x }}{{\partial y}} \approx \rho c_s
\lambda \frac{U}{\delta }$. With the symmetries in the problem, the contributions to the stress tensor of the kind
given in Eq. (\ref{stresstensor}) disappear, resulting in a second order correction \cite{landau1981} $\sigma
_{xy}^{(2)}  = \rho \lambda ^2 \frac{{\partial u_x }}{{\partial y}}{\rm{ }}(\nabla  \cdot \textit{\textbf{u}}) \approx
\rho \lambda ^2 \frac{{U^2 }}{{x\delta }}$. Thus, the expansion becomes $ \sigma _{xy}^{(1)}  + \sigma _{xy}^{(2)}
\approx \rho c_s \lambda \frac{U}{\delta }(1 + \frac{{\lambda U}}{{xc_s }}) \approx \rho c_s \lambda \frac{U}{\delta
}(1 + \frac{{\lambda ^2 }}{{\delta ^2 }})$. Note that the second order term can also be expressed as {\rm{Wi} since
$x/U = T$. This example establishes that the relevant linear dimension $\delta$ is no longer a geometric dimension of
the body but a dynamic characteristic of the flow and ${\rm{Kn}}_\delta   \equiv \lambda /\delta  \propto \sqrt
{{\rm{Wi}}} $ emerges as the scaling parameter.

\begin{figure*}
\includegraphics[trim=0mm 0mm 0mm 0mm,clip=true,scale=0.65]{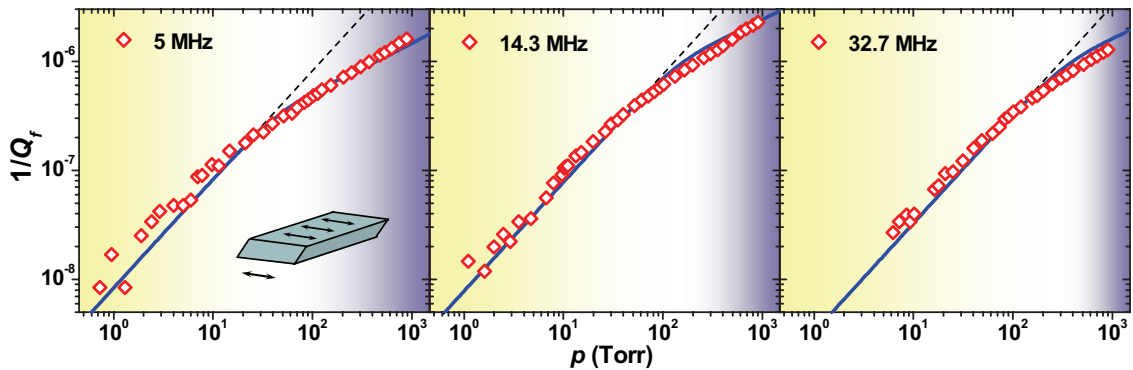}
\caption{(color online). $1/Q_f$ in quartz crystal resonators at 5 MHz, 14.3 MHz and 32.7 MHz as a function of
pressure. The inset is an illustration of the shear mode. The  $\omega \tau \approx 1$ transition from non-Newtonian
(yellow) to Newtonian (blue) flow takes place in the white regions at $p\approx$ 60, 200 and 400 Torr. The dotted lines
are asymptotes proportional to $p$. The solid lines are fits to Eq. (\ref{QfInv}) using fitting factors 0.8, 0.6 and
0.9 and $S/m=2.754$ $\rm{ m^2/kg}$, $9.827$ $\rm{ m^2/kg}$ and $6.296$ $\rm{ m^2/kg}$ (left to right).}
\label{QfInvPlot}
\end{figure*}

Even more unexpected conclusions emerge for the {\it unsteady} flow generated by an infinite plate oscillating at
(angular) frequency $\omega$ with peak velocity and amplitude $U$ and $A$, respectively (Fig. \ref{schematic}(b) and
(c)) \cite{stokes1851}. In the Newtonian limit $\omega \tau  \ll 1$, the Stokes boundary layer thickness, $\delta  =
\sqrt {\frac{{2 \nu}}{\omega }} $, is the only length scale in the problem. In this limit, ${\rm{Kn}}_\delta   \propto
\sqrt {{\rm{Wi}}}$ holds as above, since ${\rm{Kn}}_\delta   = \frac{\lambda }{\delta } \approx \sqrt {\frac{{\omega
\lambda ^2 }}{\nu}}  \approx \sqrt {\omega \tau } $. Due to its geometric simplicity, this problem can be solved in the
entire dimensionless frequency range $0 < \omega \tau  < \infty $ by a summation of the Chapman-Enskog expansion of
kinetic theory \cite{yakhot2007,karabacak2007} and non-perturbatively \cite{chen2007}. The analytic solution for the
velocity field is obtained as
\begin{equation}\label{u_x}
u_x (y) = Ue^{ - y/\delta _ -  } \cos (\omega t - y/\delta _ +  ),
\end{equation}
with two new length scales, a wavelength $\delta _ +$ and a penetration depth $\delta _ -$:
\begin{eqnarray}\label{delta}
\frac{\delta }{{\delta _ \pm  }} = (1 + \omega ^2 \tau ^2 )^{1/4} \left[\cos \left( {\frac{{\tan ^{ - 1} \omega \tau }}{2}} \right) \pm \right. \nonumber\\ \left. \sin \left( {\frac{{\tan ^{ - 1} \omega \tau }}{2}} \right) \right].
\end{eqnarray}
In the limit $\omega \tau  \gg 1$, $\delta$  disappears from the problem and $\delta_-$ becomes the relevant length
scale. However, as $\omega \tau  \to \infty $, the {\it first} Knudsen number saturates: ${\rm{Kn}}_{\delta _ -  }
\equiv \frac{\lambda }{{\delta _ -  }} \to \frac{1}{2}$, indicating that in this limit ${\rm{Kn}}_{\delta _ -  } $
cannot appear as an expansion parameter in the stress-strain relation or as a scaling parameter in Eq. (\ref{u_mod}).
The {\it second} Knudsen number, $\rm{Kn}_{\delta _ +  }  \equiv \frac{\lambda }{{\delta _ +  }} \to \omega \tau $,
becomes the exact same expansion parameter derived from the Chapman-Enskog expansion applied to the Boltzmann-BGK
equation \cite{yakhot2007}. We conclude that the only relevant scaling parameter for the oscillating plate problem
valid in both Newtonian and non-Newtonian regimes is ${\rm{Wi}} = \omega \tau$, re-emphasizing that $\rm{Wi}$ does not
contain any information about the linear dimensions of the oscillating body. Moreover, as long as $A \ll L$ and
${\mathop{\rm Re}\nolimits} \sim 0$, the flow over the oscillating body remains tangential along the surface, following
the natural curvatures.

In order to provide experimental support for the predicted universality, we studied energy dissipation in flows
generated by oscillating solid surfaces.  For a large plate, the average energy dissipated per unit time can be
obtained as $\dot E = \frac{{SU^2 }}{2}f(\omega \tau )\sqrt {\frac{1}{2}\omega \mu \rho }$ by considering the shear
stress on the plate. Here, $S$ is the surface area and $f(\omega \tau )$ is the scaling function found as
\cite{karabacak2007}
\begin{eqnarray}\label{f}
f(\omega \tau ) = \frac{1}{{(1 + \omega ^2 \tau ^2 )^{3/4} }}\left[ (1 + \omega \tau )\cos \left( {\frac{{\tan ^{ - 1} \omega \tau }}{2}} \right)\right. \nonumber\\ \left.  - (1 - \omega \tau )\sin \left( {\frac{{\tan ^{ - 1} \omega \tau }}{2}} \right) \right].
\end{eqnarray}
For a mechanical resonator with resonance frequency $\omega/2\pi$, the dissipation can be translated into a fluidic
quality factor $Q_f$ from the relation
\begin{equation}\label{QfInv}
\frac{1}{Q_f}  = \frac{{\dot E}}{{\omega E_{st} }} = \frac{S}{m}f(\omega \tau )\sqrt {\frac{{\mu \rho }}{{2\omega }}},
\end{equation}
where $E_{st}  = \frac{{mU^2 }}{2}$ is the energy stored in the resonator with mode mass $m$. In a simple ideal gas,
such as nitrogen at pressure $p$, $\tau \propto 1/p$ .  $1/Q_f$ can then be expressed as a function of $p$ and follows
two different asymptotes: $1/Q_f  \propto p$  at low $p$ (non-Newtonian) and $1/Q_f \propto p^{1/2}$
 at high $p$ (Newtonian). The transition between the asymptotes takes place at $\omega \tau \approx
1$, and shifts to higher $p$ as $\omega$ is increased since $\tau \propto 1/p$.

In our experiments, we measured  the $Q$ factor of quartz crystals, microcantilevers and nanomechanical beams in dry
nitrogen as a function of $p$ using electrical \cite{lea1984} and optical techniques \cite{karabacak2007}. We excited
the resonators at very small oscillation amplitudes ($A \ll L$) around their resonances. The energy losses arising both
from the fluid and the resonator itself (coupled to the measurement circuit) determine the overall (loaded) quality
factor as $1/Q_l=1/Q_f(p) + 1/Q_r$. At low $p$, $1/Q_f \to 0$, allowing a measurement of $1/Q_r$, and subsequently,
$1/Q_f (p)$ \cite{karabacak2007}.

\begin{figure*}
\includegraphics[trim=0mm 0mm 0mm 0mm,clip=true,scale=1.57]{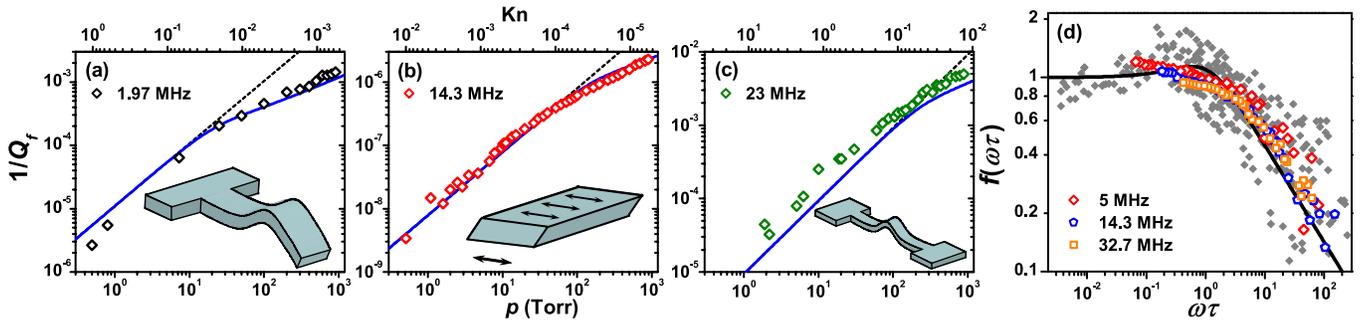}
\caption{(color online). Fluidic dissipation across length scales. (a) A microcantilever with dimensions $l\times w
\times t=$ 125 $\mu$m $\times$ 36 $\mu$m $\times$ 3.6 $\mu$m moving in its first flexural harmonic mode in the
out-of-plane direction. (b) A quartz crystal of diameter 0.5 cm and thickness 0.1 mm moving in fundamental shear mode.
(c) A nanomechanical doubly-clamped beam with $l\times w \times t=$ 17 $\mu$m $\times$ 500 nm $\times$ 280 nm moving in
fundamental out-of-plane flexural mode. The upper axes in (a)-(c) show ${\rm{Kn}} = \lambda /L$ with $L \approx \sqrt
S$. The dotted lines are asymptotes proportional to $p$ and the $\omega \tau \approx 1$ transition takes place at
$p\approx$ 20, 200 and 300 Torr. All solid lines are obtained from theory. (d) Scaling of fluidic dissipation. The
solid line shows the scaling function $f(\omega \tau )$ of Eq. (\ref{f}). The dissipation data of macroscopic quartz
resonators are collapsed using Eq. (\ref{QfInv}) with the fitting factors and $S/m$ values mentioned in Fig.
\ref{QfInvPlot}. The data points plotted in grey are from smaller flexural resonators of \cite{karabacak2007}. For the
flexural resonators, the fitting factors are 2.8; $S/m$ values are calculated from geometry and mode shape
\cite{karabacak2007}.} \label{fPlot}
\end{figure*}

In Fig. \ref{QfInvPlot}, we show the basic aspects of the universality on similar sized macroscopic quartz crystals
moving in fundamental shear mode (inset) at $\omega /2\pi  = $ 5, 14.3 and 32.7 MHz. In each plot, the slope of
$1/Q_f(p)$ changes due to the transition from non-Newtonian (yellow) to Newtonian (blue) flow.  In order to fit the
data to in Eq. (\ref{QfInv}), we used experimentally measured $S/m$ values \footnote{The resonance frequency shift due
to deposition of a gold film of known thickness was used in the Sauerbrey formula to determine $S/m$.} and the recently
suggested empirical form $\tau \approx \frac{{1.85 \times 10^{ - 6} }}{p}$ ($\tau$ is in s when $p$ is in Torr)
\cite{karabacak2007}.  The flow for the small amplitude shear mode oscillations of the quartz crystals matches the
large plate problem to a very good approximation \cite{stockbridge1966,krim1988}. The only non-ideality may come from
the velocity distribution on the surface of the resonator \cite{martin1989}, which explains the small deviation of the
fitting factors from unity.

The universality requires that the characteristics of the flow remain size and shape independent. We establish this
aspect in Fig. \ref{fPlot}(a)-(c) by comparing data on resonators, which span a broad range of linear dimension $L$ and
oscillate in different modes: a macroscopic quartz crystal in shear-mode at 14.3 MH; a microcantilever and a
nanomechanical doubly-clamped beam in flexural modes at 1.97 MHz and 23 MHz, respectively. We take the dynamically
relevant linear dimension of the flow as $L \approx \sqrt S $, determined by the surface area. For the quartz crystal,
$L_Q \approx {\rm{10}}^{{\rm{ - 2}}}$ ${\rm{m}}$, set by the electrode diameter \cite{herrscher2007,capelle1990}. For
the cantilever, $L_C \approx 10^{-4}$ $\rm{m}$ and for the  beam,  $L_B  \approx 5 \times 10^{-6}$ $\rm{m}$. The modes
are illustrated in the insets; the flow remains tangential to the solid surface along the gentle curvatures. ${\rm{Kn}}
= \lambda /L$ for the devices are shown on the upper axes. Two things are noteworthy: all curves look similar, and $L$
or $\rm{Kn}$ appears to have no effect on the flow in our parameter range \cite{verbridge2008}. Fig. \ref{fPlot}(d)
shows the dissipation data of macroscopic quartz resonators collapsed onto a dimensionless plot along with data of
smaller resonators from \cite{karabacak2007}. In analyzing the data of flexural resonators, a fitting factor of 2.8 was
used as opposed to the near unity fitting factors used for the shear-mode quartz crystals \footnote{This discrepancy is
possibly due to the difference between flexural and shear motion. In out-of-plane flexural motion, the curvature of the
surface results in increased tangential flow velocity and increased dissipation. This effect has been discussed, for
instance, for a long cylinder in \cite{landau1959} on p. 90. Oscillations perpendicular to the cylinder axis (cross
flow) results in a factor of two more dissipation than comparable parallel oscillations.}.

In our experiments, $\tau$ is determined by the microscopic interaction of gas molecules and a solid surface. By
naively treating the nitrogen as a gas of hard spheres, one can obtain $\tau  \approx \frac{{0.180 \times 10^{ - 6}
}}{p}$ ($\tau$ is in s when $p$ is in Torr).  This value, however, only reflects interactions between gas molecules, as
would happen away from surfaces in the bulk region of the gas. The observed dependence, which is roughly an order of
magnitude larger, points to the importance of gas-surface interactions.

Our data provide evidence for a transition from purely viscous (Newtonian) to viscoelastic (non-Newtonian) dynamics in
oscillating flows of simple gases in simple geometries. In our experiments, ${\mathop{\rm Re}\nolimits} \sim 0$ and
non-linear effects, such as hydrodynamic instabilities and viscoelastic turbulence, are not present. The observed
transition is due to the {\it intrinsic} dynamical response of the simple fluid to high-frequency perturbations.
Similar observations are commonplace in macroscopic flows of concentrated long-chain polymer solutions, where $\tau$
can be long and, consequently, ${\rm{Wi}} \ge 1$ due to the relatively slow polymer dynamics
\cite{Bird1987,groisman2000,magda1988}. In rheology, polymers are often treated as elastic springs, and viscoelastic
behavior of polymer solutions is attributed to the direct contribution of polymer molecules to the stress tensor. In
this sense, our work points to a deep dynamical connection between oscillating flows of complex and simple fluids
\cite{Bird1987}.

We thank N. O. Azak, M. Y. Shagam and A. Vandelay for experimental help and discussions.  This work was supported by
Boston University through a Dean's Catalyst Award and by the NSF through Grant No. CBET-0755927.

\end{document}